\documentclass[twocolumn]{aastex61}
\pdfoutput=1 
\usepackage{amsmath, amstext}
\usepackage[T1]{fontenc}
\usepackage{apjfonts} 
\usepackage[figure,figure*]{hypcap}

\usepackage{color}
\usepackage{soul}
\sethlcolor{orange}

\newcommand{\kms}	{{\rm km}\, {\rm s}^{-1}}

\shorttitle{AASTeX 6.1 Template}
\shortauthors{Van Oort et al.}

\begin{document}

\title{CASI: A Convolutional Neural Network Approach for Shell Identification}

\author{Colin M. Van Oort}
\affiliation{Complex Systems Center, University of Vermont, Burlington, VT 05405, USA}
\email{cvanoort@uvm.edu}
\author{Duo Xu}
\affiliation{Department of Astronomy, The University of Texas at Austin, Austin, TX 78712, USA}
\author{Stella S.R. Offner}
\affiliation{Department of Astronomy, The University of Texas at Austin, Austin, TX 78712, USA}
\author{Robert A. Gutermuth}
\affiliation{Department of Astronomy, University of Massachusetts - Amherst, Amherst, MA 01003, USA}

\begin{abstract}
We utilize techniques from deep learning to identify signatures of stellar feedback in simulated molecular clouds. 
Specifically, we implement a deep neural network with an architecture similar to U-Net and apply it to the problem of identifying wind-driven shells and bubbles using data from magneto-hydrodynamic simulations of turbulent molecular clouds with embedded stellar sources.
The network is applied to two tasks, dense regression and segmentation, on two varieties of data, simulated density and synthetic $^{12}$CO observations.
Our Convolutional Approach for Shell Identification ({\sc casi}) is able to obtain a true positive rate greater than 90\%, while maintaining a false positive rate of 1\%, on two segmentation tasks and also performs well on related regression tasks.
The source code for {\sc casi} is available on GitLab.
\end{abstract}


\section{Introduction}\label{intro}

Forming stars influence their environment by injecting energy over a large dynamic range with different sources contributing at different times and characteristic length scales.
Stellar feedback has been invoked to explain a host of phenomena including the relation between dense cores and the stellar Initial Mass Function \citep[IMF,][]{alves07,Offner14ppvi}, the longevity of turbulence within molecular clouds \citep{cunningham06,wang10,Offner18}, the properties of multiple star systems \citep{Offner16} and the global efficiency of star formation \citep{krumholz07, lee15,federrath15}.
Nevertheless, the energetics and impact of feedback remains poorly constrained.

Identifying feedback signatures and quantitatively disentangling the interaction with the environment are notoriously difficult.
For decades, astronomers have studied the distribution of gas in the interstellar medium by making 2D dust emission and absorption maps and 3D atomic and molecular spectral cubes.
A variety of algorithms have been developed to identify peaks in the data, namely cores and filaments, including {\sc clumpfind}, {\sc dendrograms} and {\sc getfilaments} \citep{williams94, goodman09, menshchikov13}.
However, simple structure identification algorithms like these fail to identify feedback signatures, which exhibit a variety of complex morphologies. 
Statistical approaches, such as principle component analysis and the spectral correlation function provide a means to quantify the underlying impact of feedback on the turbulent cloud structure; however, many statistics commonly applied to spectral line cubes are relatively insensitive \citep{boyden18}.
Consequently, the imprint of feedback is usually identified ``by eye" \citep{churchwell06, arce10, arce11, li15}.

The human brain is a superb tool for parsing complex images \citep{zhangzhang10}, and a variety of papers have used visual identification to study feedback in surveys of individual regions \citep[e.g.,][]{konyves07, arce10, arce11, narayana12, li15}.  Features produced by stellar winds and outflows resemble shells, bubbles or cones in intensity maps, which is one way they can be visually identified  \citep{churchwell06,simpson12,Offner11}. In spectral line data, such as CO observations,  feedback often appears connected over a range of velocities (frequencies), so astronomers often identify feedback by searching for coherent three-dimensional structures \citep{arce10,arce11,li15}.
Meanwhile, the explosion of data over the last decade and production of large surveys, such as those covering the entire Galactic plane, have outstripped the analysis capacity of professional astronomers.
This has led to a variety of ``citizen science" efforts, in which interested members of the public visually inspect and characterize the data.
Galaxy Zoo, which has undergone a number of iterations, involved millions of people, and produced dozens of papers to date, is the highest-profile of these initiatives \citep[e.g.,][]{lintott2008}.
Recently, the Milky Way Project applied the power of citizen science to the identification of stellar feedback in the Spitzer Galactic plane surveys GLIMPSE and MIPSGAL.
This effort yielded a catalog containing the locations and sizes of thousands of new bubbles in the Milky Way \citep{simpson12}.

However, human classification, while formidable, has several disadvantages.
Although numerous people devote significant time to data parsing, citizen hours are finite and only certain problems can be formulated into simple pattern searches for non-experts.
Moreover, classifications are subjective and differ between people.
This can produce different catalogs and conclusions for the same data even between experts (compare \citet{narayana12} with \citet{li15}, for example).

The very nature of feedback ensures that human identification will be ambiguous. Since stellar feedback acts on the interstellar medium, which by nature has a strongly inhomogenous density and velocity distribution, signatures are usually asymmetric and often blend into the turbulent background \citep{arce10,arce11}.
Voids, low density regions that are produced by supersonic turbulence, may also mascarade as feedback-driven bubbles, causing false positives.
Although stellar feedback can accelerate cloud gas to velocities above the mean cloud turbulent velocity, the peak velocity of the feedback is sensitive to the source orientation with respect to the line of sight and its location relative to the cloud boundary, where gas changes phase from molecular to atomic \citep{arce10,Offner11,li15}.
These complications mean that even experts have trouble unambiguously and accurately identifying feedback.

Algorithmic approaches to identifying bubbles have been utilized in order to remove some of the subjectivity of bubble identification, while also removing some of the burden from human identifiers \citep{giri2017bubble}. However, more traditional algorithmic approaches tend to lack the flexibility required for widespread application.

One alternative approach is machine learning, a sub-field of computer science in which algorithms adapt to patterns and correlations in data.
Machine learning is now a mature field, and is commonly applied to pattern recognition problems, including topics ranging from genome sequencing to face recognition to drug discovery \citep{MLreview}.
Machine learning can automate the process of feature identification, scale efficiently to large data sets, and produce repeatable catalogs.
However, to date it has been applied relatively sparsely to problems in astrophysics.

In this work, we present our convolutional approach for shell identification ({\sc CASI}) \citep{van_oort_colin_m_2019_2695533}\footnote{The source code for {\sc CASI} is available on GitLab: \href{https://gitlab.com/casi-project/casi-2d}{https://gitlab.com/casi-project/casi-2d}}.
{\sc CASI} is a convolutional neural network, a variety of artificial neural network (ANN) where the primary unit of computation is the convolution operation rather than simple matrix multiplication.
For an overview of convolution arithmetic in the context of machine learning, see Appendix \ref{appendix.nn-ops} for a brief overview and \cite{dumoulin2016guide} for a more comprehensive guide\footnote{An associated GitHub page provides helpful animations: \url{https://github.com/vdumoulin/conv_arithmetic}.}.
ANNs are a computational model that is loosely inspired by biological neural networks, where the fundamental unit of computation is a single neuron that receives one or more stimuli and provides one or more output signals.
See section 3 in \citet{lieu2019detecting} for a brief overview of ANNs that targets the astronomy audience.

{\sc CASI} is designed to identify feedback signatures in molecular clouds, with a focus on wind-driven bubbles created by intermediate-mass stars.
This is motivated by the observation that such shells identified in nearby star-forming regions, like the Perseus molecular cloud, have a huge impact on the cloud energetics and evolution \citep{arce11}.
Magneto-hydrodynamic (MHD) simulations with embedded sources are used to train our method and investigate its efficacy.

In the remainder of \S\ref{intro}, we summarize relevant machine learning applications in the literature.
We describe our method in \S\ref{method} and present results in \S\ref{results}.
Finally, conclusions and discussion are provided in \S\ref{conclusions}.

\subsection{Machine Learning for Image Tasks}\label{sec.nn-vision}

\cite{hubel1968receptive} identified specialized neurons in the visual cortices of cats and monkeys that process small, partially overlapping regions of their visual field.
This pattern of local, overlapping connectivity inspired the design of the Neocognitron \citep{fukushima1982neocognitron}, a neural network based approach to character and digit recognition.
However, difficulties encountered when training networks with more layers and a lack of sufficient training data led to a decline in the popularity of ANNs, with alternative methods such as support vector machines (SVMs) receiving more attention.

More than a decade later \cite{lecun1998gradient} introduced LeNet-5, a Convolutional Neural Network (CNN) that broke the record for character recognition performance and became a baseline architecture for many applications of CNNs that followed, contributing to a resurgence the popularity of ANNs.
This resurgence ushered in a wave of research and targeted hardware improvements that allowed ANNs to overtake many competing machine learning algorithms and attain state-of-the-art results on a variety of tasks, rivaling human performance in some cases \citep{nasser2017metrics}.

In addition to character recognition, CNNs have been successful at image classification, object detection, semantic segmentation, and image denoising/artifact removal to name a few.
For a broad overview of the techniques involved in CNNs and their applications see \cite{gu2015recent}.

\subsection{Previous Applications to Astronomical Data Analysis}\label{sec.prev}

Machine learning techniques have been applied to structure detection in astronomical data several times with varying degrees of success.

\cite{beaumont2011classifying} used SVMs to segment $^{12}$CO data containing a supernova remnant partially obscured by a molecular cloud, reaching >90\% accuracy when classifying hand-labeled pixels as belonging to the supernova remnant or molecular cloud.

SVMs are a supervised learning method that classifies data by finding a decision boundary that simultaneously minimizes classification error and maximizes the distance between the boundary and closest samples of any class.
SVMs may also be applied to regression problems.
Such applications are often referred to as support vector regression.
Since SVMs attempt to maximize the margins about the decision boundary they tend to generalize well and feature robustness to minor perturbations of input data.
Interested readers should refer to \cite{bennett2000support} for an overview of SVMs.

\cite{beaumont2014milky} developed {\sc Brut}, a method that utilizes Random Forest classifiers, to identify bubbles and similar structures in color-composite images from the Spitzer Space Telescope.
However, {\sc Brut} is sensitive to the position of the bubble in the image, making wide-field searches computationally expensive \citep{xu17}.

Deep learning is a relatively new and rapidly evolving sub-field of machine learning that features ANNs with sophisticated architectures and greater numbers of layers.
Relatively few astrophysical applications utilize deep learning techniques, which may be partly due to the age and the rapid research pace of the field.
\cite{daigle2003automatic} utilized a Multi-Layer Perceptron (MLP), a simple neural network architecture that features consecutive layers of densely connected artificial neurons, to identify expanding shells in the Canadian Galactic Plane Survey, obtaining a $0.6\%$ false positive rate.
Later \cite{daigle2007automatic} compared the performance of the MLP against two alternative network architectures, the competitive network and the growing neural gas network, on similar data.
There was no clear winner in this comparison, since all three networks were able to correctly identify 10 out of the 11 bubbles considered when evaluated using a leave-one-out cross-validation method.

\cite{dieleman2015rotation} applied a CNN to the morphological classification of annotated images from the Galaxy Zoo project, attaining an accuracy $>99\%$ for images where human annotators strongly agreed upon the classification label.
The authors suggest that a machine learning system could be used to classify the ``easy" images, leaving the more difficult cases for human annotators.
Filtering the images in this way could lead to a reduced workload for human annotators when processing large surveys.

\cite{lanusse2017cmu} trained a CNN to identify the existence of gravitational lensing in simulated data that was constructed to resemble Large Synoptic Survey Telescope (LSST) observations.
This approach reached a true positive rate $\geq 80\%$ while maintaining a false positive rate of $1\%$ on samples with varying signal to noise ratio.

The network employed in \cite{lanusse2017cmu} utilizes residual connections, a network architecture feature introduced by \cite{he2016deep} where identity connections combine the input and output data of a block of operations.
Residual connections effectively change the underlying model of a network, or network component, from $y = f(x)$ to $y = f(x) + x$, and encourage the network to learn iterative transformations of the input rather than a direct mapping \citep{jastrzebski2017residual}.
Networks and network components that incorporate residual connections can easily learn the identity function, which allows them to mitigate the effects of harmful or under-performing components during learning.
These properties allow architectures with residual connections to effectively utilize a greater number of layers and a larger number of model weights than architectures that do not include residual connections.
See Section \ref{appendix.residual} for more details on residual architectures.

\cite{primack2018deep} utilized a simple CNN to classify images from the CANDELS survey into one of three phases of galaxy evolution.
The network is trained using simulated CANDELS-like observations and then applied to real data, reaching so-called ``Blue Nugget'' phase galaxy identification accuracy of around $80\%$.
This application involves a relatively small data set, thus the authors implemented several measures to keep the network from over-fitting, including data augmentation and dropout.

\cite{lieu2019detecting} trained a CNN to classify solar system objects from other astronomical sources in simulated data.
The network is initialized with weights that were trained on the ImageNet data set and then fine-tuned on 7512 simulated Euclid images.
Similar to \cite{primack2018deep}, this work utilizes various techniques to mitigate over-fitting, including batch normalization (see Appendix \ref{appendix.nn-ops}), dropout and data augmentation.
After testing several modern CNN architectures, \cite{lieu2019detecting} are able to reach an accuracy of $95.6\%$ when distinguishing between four classes of stellar objects.

Most recently, \cite{diaz2019classifying} investigated the classification of simulated galaxies into three classes.
CNNs were applied to this task, using data generated from N-body simulations as training data, and they were able to obtain an accuracy exceeding $99\%$.

The extent of previous work in this area, as well as the lack of a comprehensive and automated solution, motivates further application of machine learning techniques to structure identification in studies of star formation and the interstellar medium.
In this study we apply the U-Net architecture, which is described in the following section, to several tasks derived from MHD simulation data.

\section{Method Overview} \label{method}

\subsection{Neural Network Architecture}\label{sec.architecture}
\begin{figure*}
\centering
\includegraphics[width=\textwidth]{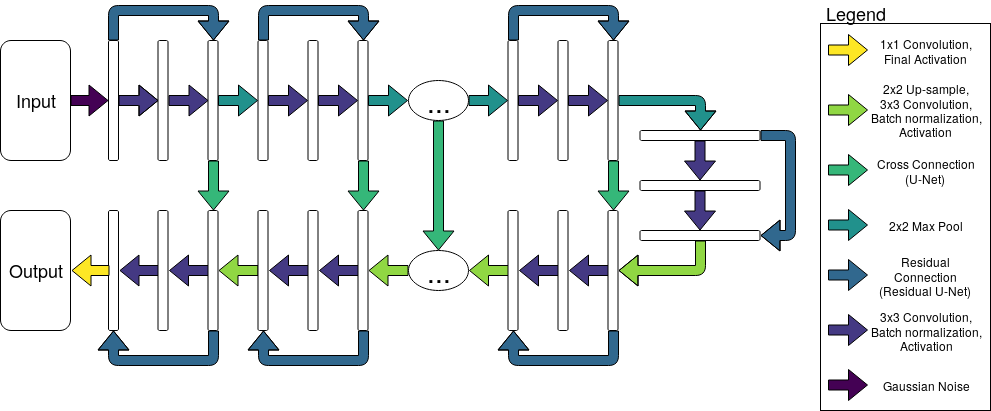}
\caption{
A flow-chart style depiction of our Residual U-Net architecture, which maps a 4D stack of images with dimensions (number of images, height of images, width of images, number of image channels) to a 4D stack of images (number of images, height of images, width of images, number of output channels).
In the context of astronomy data, the ``images" in question may be slices of observational data volumes and the height/width are literally the height and width of the observational data (in pixels/voxels). In this work we only utilize a single input image channel (gas density in a voxel, CO intensity in a voxel, etc.), however, it is possible to supply the network with multiple varieties of data using multiple input image channels.
The first half of the network uses down-sampling operations to compress input features and construct higher-level representations, while the second half of the network uses up-sampling operations to reconstitute the abstract representations.
Cross connections allow information from the down-sampling path to be utilized in the up-sampling path, leading to mappings that benefit from the combination of coarse and fine grained features.
The ellipsis-within-oval graphics indicate that the depth of the architecture is variable and may be modified by the user.
}
\label{fig.compress_arch}
\end{figure*}
In this work we employ a Residual U-Net, a variant of the U-Net architecture developed by \cite{ronneberger2015u} where the fundamental unit of construction is a residual block \citep{he2016deep}, rather than a single convolution.
A residual block is simply a sequence of consecutively applied convolution operations that are spanned by a residual connection. See Appendix \ref{appendix.residual} for more details.

The U-Net architecture and its derivatives have grown in popularity since their introduction, and Residual U-Nets in particular have been applied to a wide variety of problems including road segmentation \citep{zhang2018road}, detection of pulmonary nodules \citep{lan2018run}, segmentation of optic nerve tissue \citep{devalla2018drunet}, and several other medical segmentation tasks \citep{zhu2018anatomynet}.

Figure \ref{fig.compress_arch} displays our Residual U-Net architecture and provides details on the structure of each sub-component.
Beyond the addition of residual connections, we also make a few other small alterations to the original U-Net architecture.

In particular, we utilize padded convolutions\footnote{Padded convolutions augment the convolution operation by extending the spatial dimensions, e.g. height and width, of the input with generated data. One common padding scheme is to apply a band of zeros that is half as wide as the spatial dimensions of the convolution filter in that direction. This scheme results in a convolution whose input volume and output volume have identical dimensions when the convolution filters have odd spatial dimensions (e.g. 3, 5, 7, ...).} in place of unpadded convolutions, which results in feature maps with identical spatial dimensions at corresponding levels of the down-sampling and up-sampling paths.
This removes the need to apply cropping to the cross connections and allows the depth of the network to be modified more easily when a particular problem benefits from the use of higher-level features.
We make use of batch normalization prior to each activation function, which was not used by \cite{ronneberger2015u}, since it can stabilize training and act as a light regularizer \citep{ioffe2015batch}.
Note that batch normalization is not strictly necessary and may have a negative effect on performance for some tasks, thus it may be useful to re-evaluate its use when applying this architecture to new problem domains.

\subsection{Training}\label{sec.training}
We utilize stochastic gradient descent (SGD) with momentum to train our networks, following results from \cite{wilson2017marginal} that suggest SGD may provide better generalization properties than adaptive step size methods, such as ADAGRAD \citep{duchi2011adaptive} and ADAM \citep{kingma2014adam}.
\cite{ruder2016overview} provides an excellent overview of gradient descent algorithms, with a focus on variants used in deep learning research and applications.

SGD is an optimization algorithm where the parameters of a function, such as the weights of a neural network, are adjusted using the gradient of a loss function with respect to those parameters.
The loss function provides a performance criterion and the gradient of the loss with respect to the model parameters indicates how the parameters should be adjusted in order to reduce the loss.
The backpropagation algorithm, an application of the chain rule from differential calculus, distributes the gradients backwards through the network starting from the final layer.

The behavior of SGD can be controlled via the use of several parameters including the learning rate, batch size, and momentum intensity.
The learning rate scales the magnitude of weight updates applied to the network in each step of SGD.
Utilizing learning rates that are too high can lead to divergence, where the loss increases after each update and the network fails to learn, while learning rates that are too low may lead to premature convergence and extended training times.

Batch size determines how many training samples will be used to calculate the gradient at each step of SGD.
Utilizing a batch size of one results in what is usually referred to as online SGD, while a batch size equal to the size of the training set results in batch SGD, and the use of batch sizes that fall between these two extremes results in mini-batch SGD.
The batch size parameter features a trade-off between calculation speed and gradient accuracy when considering smaller vs.\ larger batch sizes.
Nearly all modern applications of ANNs use mini-batch SGD for training, since online SGD can introduce too much noise into the training process and batch SGD tends to take too long to converge, though there is not a strong consensus on the optimal batch size setting.
\cite{masters2018revisiting} indicate that smaller batch sizes, between 2 and 32, tend to work well in many cases, on the other hand, \cite{hoffer2017train} suggest that larger batch sizes may also be effective, as long as the training duration is extended accordingly.

Momentum is an extension to SGD where each weight update is a linear combination of the current gradient and the previous weight update, which can reduce oscillations in weight updates and speed up training convergence \citep{goh2017why}.
The momentum intensity parameter usually falls in the range $[0, 1)$ and controls the fraction of each weight updated that comes from the previous update.
For example, setting the momentum intensity to $0.9$ will result in each update consisting of the previous update multiplied by $0.9$ plus the current gradient multiplied by $0.1$.
Alternatively, the momentum intensity may simply act as a learning rate applied to the previous weight update, rather than also scaling the current update.

We train our networks for 200 epochs\footnote{
A single training epoch involves several gradient updates, such that the network is exposed to each sample of the training set exactly once.
In our case a single epoch consists of $\lceil N_T / B \rceil$ gradient updates, where $N_T$ is the number of training samples and $B$ is the batch size.} of SGD with the momentum parameter set to 0.9 and a batch size of 8.

\begin{figure}
\centering
\includegraphics[trim=1.1cm 0.25cm 1.25cm 0.8cm, clip, width=\columnwidth]{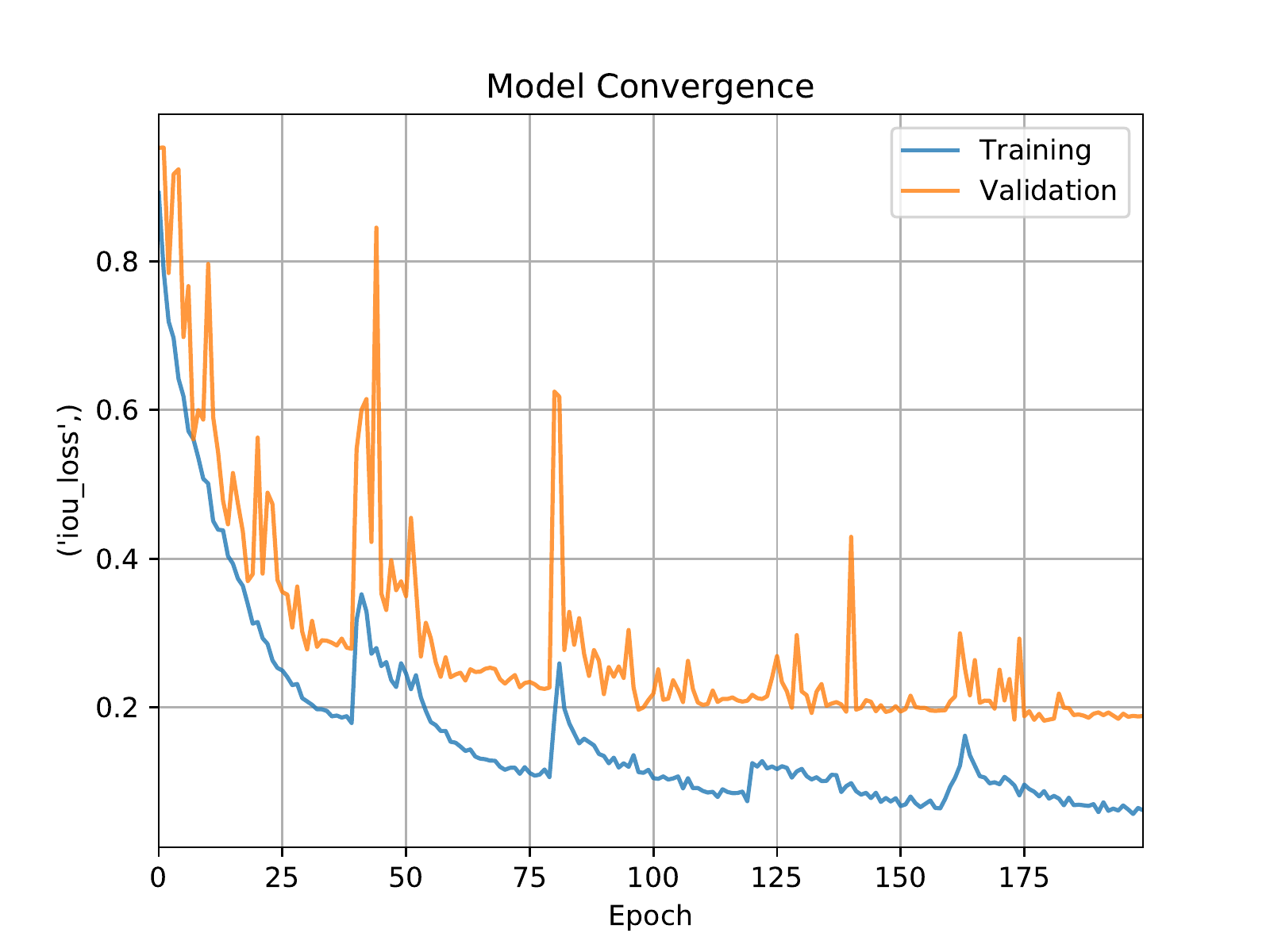}
\caption{An example of the training and validation loss curves for a Residual U-Net trained on the CO segmentation task using the Intersection over Union loss function. The spikes that occur every 40 epochs are a feature of the cyclic learning rate schedule. Note that by the 200th epoch, network performance appears to have converged, with little improvement in the validation loss for 50 to 75 epochs.}
\label{fig:classify_co_errors}
\end{figure}

The networks are initialized with random weights using the Glorot initialization scheme \citep{glorot2010understanding}.
We utilize the uniform distribution variant of this scheme, which draws samples from a uniform distribution over the interval [-x, x], where $$x = \sqrt{6 / (fan\_in + fan\_out)},$$
$fan\_in$ is the number of input units for a weight tensor, and $fan\_out$ is the number of output units.
Note, fewer training iterations may be required if the models are initialized with weights that have been previously trained on a similar data set and task, a process that is usually referred to as transfer learning \citep{pan2010survey}.
During training the learning rate is adjusted using the cyclic learning rate schedule described in \cite{huang2017snapshot}, with a maximum learning rate of 0.2 and 5 cycles of 40 epochs.
Additionally, the training samples are shuffled at the end of each epoch, which effectively adds a small amount of noise to the gradient updates and can reduce the chance of getting stuck in a local optimum.
Finally, the model state is saved, via a check-pointing utility, each time a new minimum error is observed on the validation set.

\subsection{Hyper-parameters and Hyper-parameter Selection}

\begin{deluxetable*}{llcr}
\tablecolumns{10}
\renewcommand{\tabcolsep}{0.07cm}
\tablecaption{Network Hyper-Parameters \label{param}}
\tablehead{
\colhead{Name} &  
\colhead{Definition} &
\colhead{Symbol} &
\colhead{Value}
}
\startdata
Batch Size        & Samples provided during each training iteration.  & $B$      & 8 \\ \hline
Depth             & Number of blocks used in network construction.    & $D$      & 4 \\ \hline
Filter Count      & Filters allotted for each convolution operation.  & $F$      & 16 \\ \hline
Noise Strength    & Std. dev. of the noise applied to network inputs. & $\sigma$ & 0.003 \\ \hline
\enddata
\end{deluxetable*}

This section provides a detailed description of relevant hyper-parameters and how they influence performance of the model discussed in Section \ref{sec.architecture}.
Table \ref{param} provides a brief summary of these hyper-parameters and the values utilized in our experiments.

The batch size of the network, which controls how many images are provided to the model simultaneously during training and inference, is determined by $B$.
As mentioned in Section \ref{sec.training}, there are trade-offs to be considered when selecting the batch size parameter.
Larger batch sizes allow samples to be processed in parallel and may reduce training and inference times at the cost of additional memory overhead.
Batch sizes greater than one allow for the aggregation of gradients over several data samples, providing more accurate gradient estimates and potentially reducing the number of training iterations required for the loss function to converge.
Small batch sizes have been shown to have a beneficial regularizing effect on deep neural networks that may improve generalization \citep{keskar2016large, hoffer2017train, masters2018revisiting}.
Though progress has been made towards improving the effectiveness of networks trained with large batch sizes \citep{hoffer2017train, smith2017dont}, we tended to use small batch sizes due to memory limitations of graphics processing units (GPUs) used to accelerate training.

The depth parameter, $D$, determines the number of fundamental blocks that are used in the construction of a particular network.
For the Residual U-Net this is the number of convolution blocks, e.g. pairs of convolutions and associated operations such as batch normalization and residual connections, present in both the compressive and decompressive paths.

Each block in the Residual U-Net contains a spatial resampling operation, max pooling in the compressive path and nearest-neighbor upsampling in the decompressive path.
See Section \ref{appendix.pooling} for a brief overview of the mechanics and benefits of max pooling.

Thus, $D$ governs the amount of dimension manipulation present in these architectures as well as the ability of the network to interact with the data at different spatial resolutions.

The depth parameter also contributes to the expressiveness of the network since each fundamental block includes one or more convolution operations.
The expressiveness of a particular network refers to its ability to accurately approximate various functions.
When considering two competing networks, $X$ and $Y$, network $X$ is more expressive than network $Y$ if the set of functions that $X$ is able to accurately approximate is a super-set of the set of functions that $Y$ is able to accurately approximate.
With this in mind, increasing $D$, and thus the number of model parameters, tends to improve the ability of the model to approximate functions and therefore increases its expressiveness.

The number of filters, $F$, indicates how many filters are allotted for each convolution operation, see Section \ref{appendix.conv} for more information about how the filters are used in the model.
Each down-sampling operation increases the number of filters allotted to down-stream convolution operations by a factor proportional to the dimension reduction, and similarly, each up-sampling operation decreases the number of filters provided for down-stream convolutions in proportion to the increase in spatial dimensions.

Additive Gaussian noise may be applied to the network inputs during training in order to avoid over-fitting, and the standard deviation of this noise is controlled by the $\sigma$ parameter.
The application of random noise to training samples can improve the robustness of the resulting method to small data perturbations and reduce the chances of over-fitting to the training data.

ANNs often require a computationally expensive hyper-parameter search process in order to reach desired performance levels.
Some factors that contribute to the computational cost of this search are the number of hyper-parameters to be optimized, resources required to train the network, and complex non-linear relationships between various hyper-parameters and final model performance.
We did not utilize a comprehensive hyper-parameter optimization method in this work, since hand-tuning alone provided adequate performance to demonstrate the effectiveness and flexibility of the method.
Instead, we refer interested readers to relevant literature.

The simplest, and arguably least efficient, hyper-parameter optimization algorithms are grid-search and random-search.
\cite{bergstra2012random} investigates the relationship between these two methods and suggests that random-search may be a better choice.

Bayesian methods may offer a more intelligent method for exploring the space of network hyper-parameters, leading to a lower computational cost.
Bayesian methods are generally more complicated than random-search or grid search, thus there is a trade-off between compute time spent on the optimization and human time spent implementing more advance methods.
\cite{snoek2012practical} provides an overview of Bayesian parameter optimization in the context of machine learning.

Evolutionary algorithms have also been successfully applied to hyper-parameter tuning.
Examples include optimization of CNN hyper-parameters with a simple population-based evolutionary algorithm \citep{bardenet2013collaborative}, optimization of SVM hyper-parameters with particle swarm optimization \citep{guo2008novel}, and optimization of ANN hyper-parameters with co-variance matrix adaptation evolution strategies \citep[CMA-ES,][]{loshchilov2016cma}.

\section{Validation} \label{results}

\subsection{Simulation Training Set}

Our study uses outputs from the simulations presented in \citet{Offner15} as a training set.
These calculations are performed with the {\sc orion2} adaptive mesh refinement (AMR) code and follow the evolution of a 5~pc turbulent piece of a molecular cloud with five randomly distributed embedded sources.
The stellar sources are represented by sink particles coupled to a sub-grid model for isotropic main-sequence stellar winds.
See \citet{Offner15} for additional details.

As a training set, the simulations have one essential advantage over observational data: they have complete information, including density, velocity, gas temperature and magnetic field at every point in the 3D volume.
Of particular importance {\sc orion2} has the capability to ``tag" the gas launched in winds and follow its progress across the domain \citep[e.g.,][]{Offner17, Offner18}.
The wind tracer field is a passive scalar, advected with the gas density, which tracks the amount of wind material in each cell. This field allows us to distinguish wind material from pristine cloud material and provides an exact map of the shells and bubbles created by the feedback (see \S\ref{sec:gastraining}). 

For our training sets, we adopt outputs at different times from simulations with two different stellar distributions and two different initial magnetic field strengths as listed in Table \ref{trainingset}. 
For training, we use only the 256$^3$ basegrid, thereby neglecting the information at higher ``adaptive" resolution.
This corresponds to a spatial resolution of $\sim 0.02$ pc.

\subsection{Gas Density Training Set}\label{sec:gastraining}

We train our method with two different types of data.
The first training set is constructed using the simulation gas density, $\rho$.
We define the wind fraction as $f_t = \rho_t / \rho$, where $\rho_t$ is the density of the wind material as tracked by the tracer field.
Pixels with values of $f_t >0.02$ are considered to be part of the feedback \citep[e.g.,][]{Offner18}.
These pixels define the target regions to be identified during training, testing and validation.

Due to the high expansion velocity of the wind shells, $v \gtrsim 1 $ km~s$^{-1}$, little mixing occurs outside the boundary of the swept-up material.
Consequently, the feedback signatures are roughly spherical but are modulated by local density and magnetic field variations.
Thus, in 2D image slices, the target training regions resemble irregular bubbles.

\subsection{Synthetic CO Emission Training Set}

The second training set is constructed from a suite of synthetic molecular line observations.
We post-process each simulation output using the radiative transfer code {\sc radmc3d}\footnote{http://www.ita.uni-heidelberg.de/~dullemond/software/radmc-3d/} to obtain a spectral cube for the $^{12}$CO(1-0) emission line.
Following \citet{Offner15}, we adopt the Large Velocity Gradient (LVG) approximation, which calculates the level populations by solving the equations for local radiative statistical equilibrium.
We use the gas densities and velocities on the $256^3$ basegrid as inputs, where we convert from total mass density to molecular number density using $n_{\rm H_2} = \rho/(2.8 m_p)$ and [$^{12}$CO/H$_2$] =$10^{-4}$ \citep{frerking82}.
Gas with temperatures exceeding 1000 K or with $n_{\rm H_2} < 50$ cm$^{-3}$, where all of the CO is likely dissociated, are assigned a CO abundance of 0.
In addition, CO molecules freeze-out onto dust grains in cold gas with densities $n_{\rm H_2} > 10^4$ cm$^{-3}$, 

and CO molecules are dissociated by strong shocks, e.g., where the gas velocity exceeds 10 km/s, so we also set the CO abundance to 0 in these regions. 
We include turbulent line broadening below the grid resolution by adding a constant micro-turbulence of 0.25 $\kms$, which is consistent with the linewidth-size relation on this scale \citep{larson1981}. 
The spectral cube resolution is $\Delta v = 0.156~ \kms$.

The tracer field, which tracks the stellar winds, records the amount of wind material in a given voxel (3D pixel).
In order to use these data to define the positive and negative detections, we combine it with the gas velocity information and construct a spectral cube (position-position-velocity) that complements the synthetic CO emission.
The approach we adopt is to map the tracer field to a density regime where $50 < n_{\rm H_2} < 10^4$ cm$^{-3}$. 
We then carry out the radiative post-processing described above.
The emitting regions in these cubes provide a map of the location of the wind-driven shells.

To account for observational resolution, we place each cube at a distance of 250 and 500 pc and convolve it with a 46'' beam, which is the resolution of the COMPLETE $^{12}$CO (1-0) survey of Perseus \cite[e.g.,][]{ridge06}.
We also add random noise assuming $\sigma_{\rm rms}= 0.15$ K, which is comparable to the noise in the COMPLETE data.

\begin{deluxetable}{lccc}
\tablecolumns{3}
\tablecaption{Model Properties\tablenotemark{a} \label{trainingset}}
\tablehead{ \colhead{Model } &  
\colhead{$t_{\rm run}$(Myr)} &
\colhead{$\dot M_{\rm tot}(10^{-6 }M_\odot {\rm yr}^{-1})$} &
\colhead{$B$($\mu G$)} 
}
\startdata   
W1\_T2\_0    & 0.0 &  0   & 13.5\\  
W1\_T2\_0.1  & 0.1 & 41.7 & 13.5\\ 
W1\_T2\_0.2  & 0.2 & 41.7 & 13.5\\ 
W2\_T2\_0.1  & 0.1 & 4.5  & 13.5\\  
W2\_T2\_0.2  & 0.2 & 4.5  & 13.5\\
W2\_T3\_0.1  & 0.1 & 4.5  & 5.6\\
\enddata
\tablenotetext{a}{Model name, output time and the total stellar mass-loss rate.
All models have $L=5$pc, $M=3762 M_\odot$, initial gas temperature $T_i=10$K, $N_*=5$.
The calculations are first evolved without sources for two Mach crossing times to allow initial cloud turbulence to develop.}
\end{deluxetable}
\vspace{0.2in}

\begin{figure*}
\centering
\includegraphics[width=2in,trim={1in .4in 1in .5in}, clip]{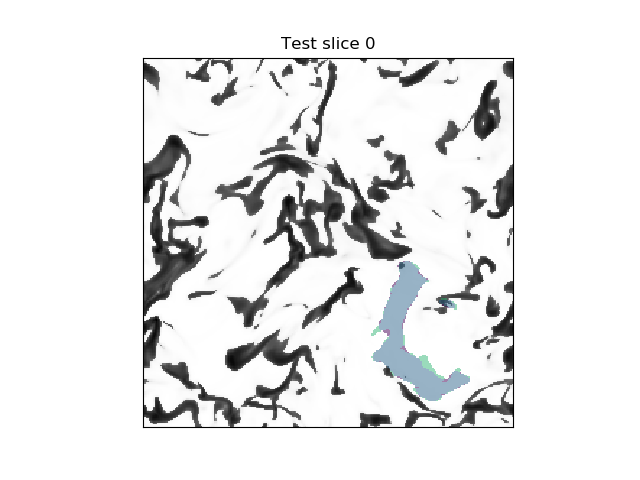}
\includegraphics[width=2in,trim={1in .4in 1in .5in}, clip]{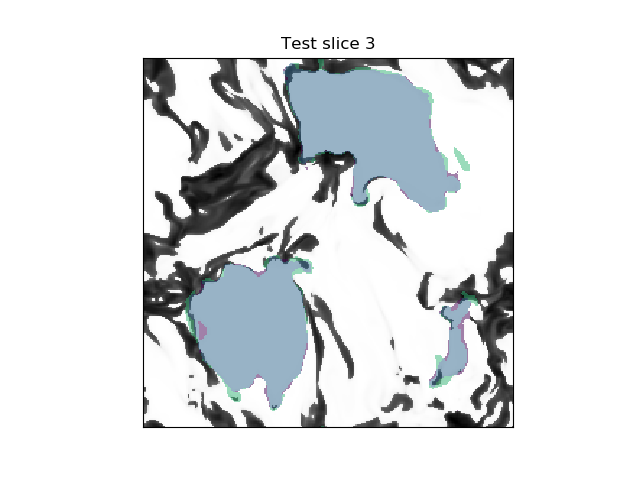}
\includegraphics[width=2in,trim={1in .4in 1in .5in}, clip]{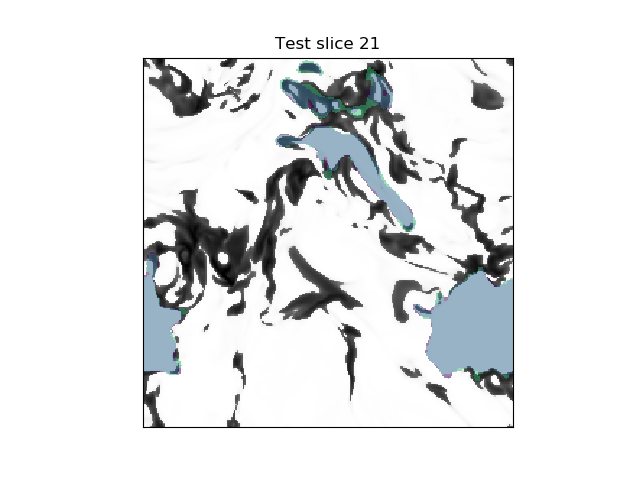}\\
\vspace{0.1in}
\includegraphics[width=2in,trim={1in .4in 1in .5in}, clip]{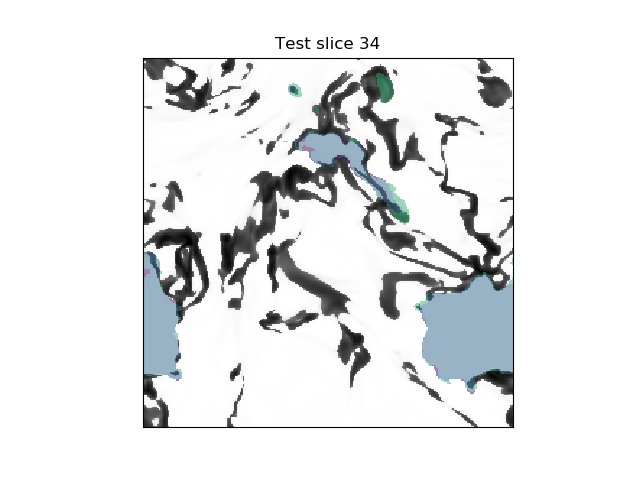}
\includegraphics[width=2in,trim={1in .4in 1in .5in}, clip]{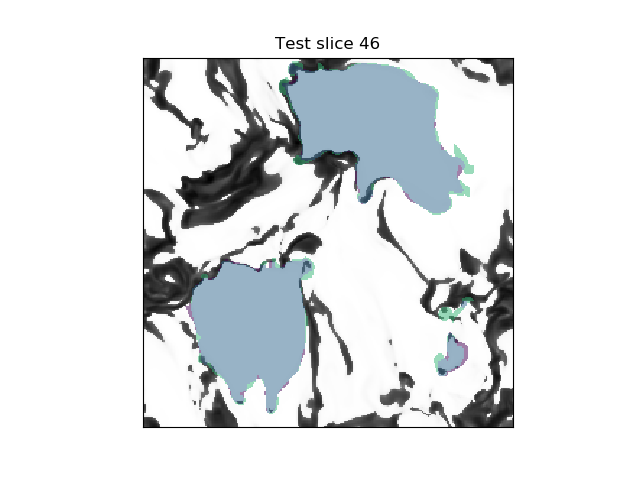}
\includegraphics[width=2in,trim={1in .4in 1in .5in}, clip]{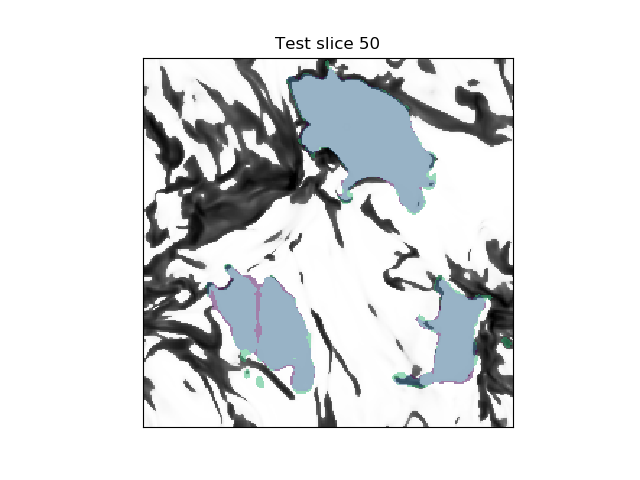}
\caption{
Density segmentation predictions from a Residual U-Net on samples randomly selected from the test set.
Each frame contains a single slice from a simulated density cube, which is shown in gray scale.
Since each slice is taken from a position-position-position cube, the x- and y-axis of each frame represent spatial coordinates within the cube.
The tiles presented here have a side length of 5~pc that is inherited from the simulation.
In each frame, true positives are shown in blue, false positives are shown in purple, false negatives are shown in green, and true negatives are not displayed.
As a pre-processing step the density data was normalized so that it is now unit-less and falls approximately in the range [-0.4, 190], where lower density regions correspond with lighter colors and higher density regions correspond with darker colors.
The color scale for the density data is identical across all tiles, and a logarithmic transformation is utilized in order to improve contrast.
\label{fig.density_segmentation}
}
\end{figure*}

\subsection{Performance Metrics}
The prediction of gas density and CO emission can be phrased in at least two ways, regression and segmentation.
In the regression phrasing, the network is expected to output a floating point value corresponding with some measure of bubble material present in each pixel (e.g., molecular line emission or a continuum map, depending on the training data set).
In the segmentation phrasing, the network is expected to classify each pixel as containing a `low' or `high' amount of bubble material.

The regression phrasing may provide more detail about perceived structures, allowing for certain kinds of analysis, such as the measurement of total bubble mass independently from the non-bubble gas along the line of sight.
However, regression methods will need to handle heavy-tailed distributions of input and output values, which could lead to poor performance.
The segmentation phrasing removes the potential difficulty of learning a heavy-tailed output distribution, but in doing so, loses some of the detail provided by regression methods.

\subsubsection{Segmentation}\label{sec.segmentation}
Segmentation masks provide less detail when compared with regressed values, but they may be more useful for identifying interesting or important regions of the input data.
For example, the outputs of segmentation models can be used to augment human efforts in processing large surveys by highlighting regions of interest or filtering regions without structures of interest.

The segmentation phrasing is achieved by selecting a threshold value, which may then be used to discretize the density and CO emission data.
The threshold value may be selected arbitrarily by the user or by using some sort of heuristic, such as selecting a certain portion of the range of the data to constitute the negative and positive classes (e.g., the lower 1\% of the range is the negative class and the upper 99\% of the range is the positive class).
We utilize a 1\% threshold since it closely aligns with features that may be visually identified.

The loss function used in the segmentation phrasing is based on the Intersection over Union (IoU) score, also known as the Jaccard Index, and is defined as
\begin{align*}
\text{IoU}\left( y,\ y' \right) = \dfrac{\text{TP}\left( y,\ y' \right)}{\text{TP}\left( y,\ y' \right) + \text{FP}\left( y,\ y' \right)},
\end{align*}
where $\text{TP}(y,\ y')$ counts the number of true positives in prediction $y'$ using the training label $y$ and $\text{FP}(y,\ y')$ counts the number of false positives.

The IoU score traditionally operates on binary inputs and is non-differentiable.
In order to facilitate the training of neural networks via gradient descent, the following differentiable approximation is used,
\begin{align*}
\text{IoU}\left( y,\ y' \right) = \dfrac{\sum_{i = 1}^{N} y[i] \cdot y'[i]}{\sum_{i = 1}^{N} y[i] + y'[i] - \sum_{i = 1}^{N} y[i] \cdot y'[i]},
\end{align*}
where $N$ is the number of pixels in $y$ and $y[i]$ is the $i$th element of $y$.
The IoU loss is simply $1 - \text{IoU}(y, y')$.

Trained models are evaluated using tools from binary classification, namely confusion matrices and derived statistics, such as accuracy, F1 Score, and Matthew's Correlation Coefficient \citep{powers2011evaluation}.

Given a confusion matrix with a number of true positives, \texttt{TP}, a number of true negatives, \texttt{TN}, a number of false positives, \texttt{FP}, and a number of false negatives, \texttt{FN}, accuracy is calculated as $$\texttt{Accuracy} = \dfrac{\texttt{TP} + \texttt{TN}}{\texttt{TP} + \texttt{TN} + \texttt{FP} + \texttt{FN}},$$
the F1 Score is calculated as $$\texttt{F1} = \dfrac{2 \times \texttt{TP}}{2 \times \texttt{TP} + \texttt{FP} + \texttt{FN}},$$
and Matthew's Correlation Coefficient is calculated as $$\texttt{MCC} = \dfrac{\texttt{TP} \times \texttt{TN} - \texttt{FP} \times \texttt{FN}}{\sqrt{(\texttt{TP} + \texttt{FP}) \times (\texttt{TP} + \texttt{FN}) \times (\texttt{TN} + \texttt{FP}) \times (\texttt{TN} + \texttt{FN})}}.$$

Receiver Operating Characteristic (ROC) curves are generated by plotting the true positive rate against the false positive rate of a model at different prediction threshold values and can provide more information about the predictive behavior of a classifier than single number statistics.

\begin{table*}
\centering
\begin{tabular}{|l|rrrrrrr|}
\hline
& True Positive &  True Negative & False Positive & False Negative & Accuracy & F1-Score & Matthews Corr. \\ \hline
Mean   &         10.82 &          87.76 &           0.47 &           0.94 &    98.59 &    91.71 &          91.07 \\
Std.   &          7.04 &           7.75 &           0.34 &           0.82 &     1.08 &    10.83 &          10.42 \\
Min.   &          0.00 &          72.49 &         < 0.01 &           0.00 &    94.58 &     0.00 &           0.00 \\
25\%   &          5.43 &          80.00 &           0.02 &           0.36 &    97.98 &    90.99 &          90.18 \\
50\%   &          9.77 &          89.17 &           0.37 &           0.72 &    98.85 &    94.82 &          93.77 \\
75\%   &         17.09 &          93.13 &           0.69 &           1.22 &    99.37 &    96.11 &          95.45 \\
Max.   &         25.44 &          99.97 &           1.52 &           3.90 &    99.98 &    98.29 &          98.11 \\ \hline
\end{tabular}
\caption{
Confusion matrix statistics for a Residual U-Net trained on the density segmentation task, computed over a test set containing 154 samples.
True positives, true negatives, false positives, and false negatives are presented as a fraction of image pixels, thus assuming values between 0 and 100.
The other three statistics, accuracy, F1-score, and Matthew's correlation coefficient, also assume values between 0 and 100, with higher values indicating better model performance.
The minimum values observed in the F1-score and Matthew's correlation coefficient are caused by a few samples with no positively labeled pixels.
}
\label{table.classify_density}
\end{table*}

\subsubsection{Regression}\label{sec.regression}
In the regression phrasing, models are trained using target values that have not been thresholded and the mean squared error (MSE) is used as the loss function.

Evaluation of regression models is traditionally dominated by the analysis of residuals, with the assumption that models featuring residuals that are tightly and symmetrically distributed about zero are better.
We utilize the following scoring function in order quantitatively evaluate and compare models according to these assumptions,
$$
f(R) = -|\langle R \rangle| - \texttt{std}(R) - |\texttt{skew}(R)|,
$$
where $f$ denotes the fitness function and $R$ denotes the computed residuals.
Note that the first term directly penalizes residual distributions whose mean value strays from 0, the second penalizes residual distributions that feature a non-zero standard deviation, and the final term penalizes residual distributions with non-zero skew.

Additional qualitative evaluation of the residuals can be obtained using histograms, kernel density estimates (KDE), and scatter plots, each providing a slightly different perspective on the distribution of residuals.

\subsection{Case Study 1: Gas Density}
In both problem phrasings the network is provided 2D slices of a 3D molecular gas density cube as input, though the expected output differs.
As noted in Sections \ref{sec.segmentation} and \ref{sec.regression}, the expected output for the regression task is the fraction of gas density associated with wind-swept bubbles and the expected output for the segmentation task is a binary mask that identifies regions with ``high'' levels of gas density associated with wind-swept bubbles.

We cut each 3D simulated density cube along its primary axes in order to form a stack of 2D slices, which are then divided into training, validation, and testing sets.
We then normalize each set of 2D slices by subtracting the mean value and dividing by the standard deviation, after which it is ready to be used in training.

\subsubsection{Density Segmentation} \label{sec:density-seg}
In Figure \ref{fig.density_segmentation}, we display examples of Residual U-Net predictions on several samples randomly selected from an unseen test set.
In the figure, the gray scale components depict re-scaled density values and and the colored components depict network predictions and errors.
Qualitatively, the model appears to correctly segment all major contiguous structures, though there may be some smaller structures that are missed.
Additionally, note that the majority of errors are located on or near the edge of identified structures and would have little effect on whether or not a particular structure is identified.
Finally, note that the upper left tile contains bubble structure that was correctly identified by \textsc{CASI} but may be difficult for a human to identify due to a lack of corresponding features in the density data.

\begin{figure}
\centering
\includegraphics[width=3.3in,trim={0in 0in 0in .55in}, clip]{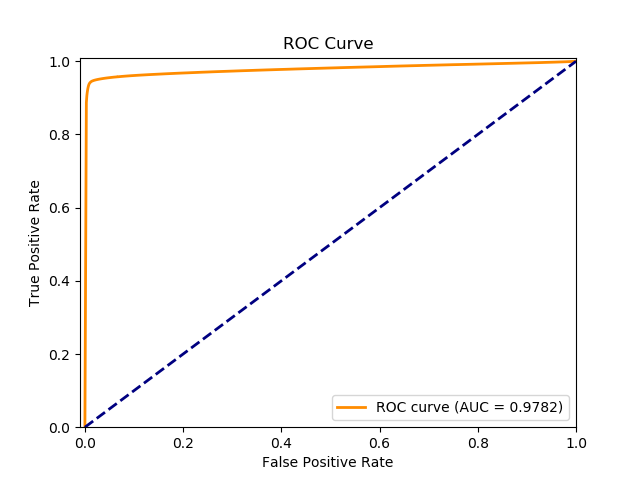}
\caption{Example ROC curve for a Residual U-Net trained on the Density segmentation task.
The dashed blue line represents $y=x$, which corresponds with the expected performance of a random binary classifier.
A true positive rate of 95.52\% is obtainable with a false positive rate of 1\%, suggesting that this method may perform well as a content filter.
\label{fig.classify_density_roc}
}
\end{figure}

In Figure \ref{fig.classify_density_roc} we present a ROC curve for the same model, which shows that the model attains a true positive rate of 95.52\% while maintaining a false positive rate of only 1\%.
Supporting this, we summarize the distributions of several binary classification statistics in Table \ref{table.classify_density}, where the classification statistics are computed across a test set of 154 samples.
In particular, Table \ref{table.classify_density} clearly highlights the low error rate obtained by our model, where the maximum fraction of false positives is 1.52\% and the maximum fraction of false negatives if 3.9\%.

In order to better grasp the effect of random initialization on final model performance we trained 60 instances of the model using the same data and parameter settings, recorded the Receiver Operating Characteristic Area Under Curve (ROC AUC) statistic for each model, and then constructed confidence intervals for the mean of the ROC AUC distribution.
The results of this experiment are presented in the first column of Table \ref{table.segmentation_scores}, which shows that CASI is able to consistently obtain ROC AUC scores close to the maximum value of 1.0.

\begin{table*}
\centering
\begin{tabular}{|l l|rr|}
\hline
Performance Stat. & Distribution Stat. & Density Segmentation & $^{12}$CO Segmentation \\ \hline
ROC AUC           & Mean               & 0.9768               & 0.909                  \\
          & Std. Error         & 0.0016               & 0.0018                 \\
          & 85\% Conf. Int.    & (0.9745, 0.9792)     & (0.9063, 0.9117)       \\ \hline
\end{tabular}
\caption{
Segmentation task performance statistics collected by training and evaluating 60 randomly initialized networks on the same training, validation, and testing splits.
The first column indicates a statistic that was computed using the predictions of each trained network, while the second column indicates a statistic that was applied to the results of the column one statistic.
The Receiver Operating Characteristic Area Under Curve (ROC AUC) statistic is calculated by computing the integral of the ROC curve, such as Figures \ref{fig.classify_co_roc} and \ref{fig.classify_density_roc}.
}
\label{table.segmentation_scores}
\end{table*}

\subsubsection{Density Regression}
\begin{figure}
\centering
\includegraphics[width=3.5in,trim={0in 0.1in 0.45in 0.55in}, clip]{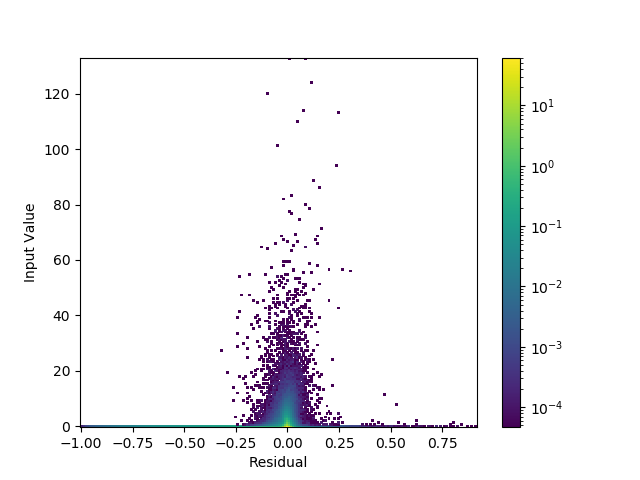}
\caption{
A 2D histogram investigating the scaling of residuals with respect to the input value for a Residual U-Net trained on the density regression task.
The color scale is logarithmic in order to increase contrast and represents the density of points associated with each residual value-input value pair.
Recall that the input values here are density values that have been scaled to have zero mean and unit standard deviation, thus the y-axis of this plot is unit-less.
Due to the heavy-tailed nature of the input values, this re-scaling results in the data that falls approximately within the range [-0.4, 190].
\label{fig.regress_density_residual_scaling}
}
\end{figure}

Applying a Residual U-Net to the density regression task leads to a tight distribution of residuals that is not strongly correlated with the size of the input value, indicating that the model has captured much of the relationship between the input and output.
Figure \ref{fig.regress_density_residual_scaling} displays a 2D histogram that shows the relationship between residuals and input values.

\begin{figure*}
\centering
\includegraphics[width=2in,trim={1in .4in .6in .55in}, clip]{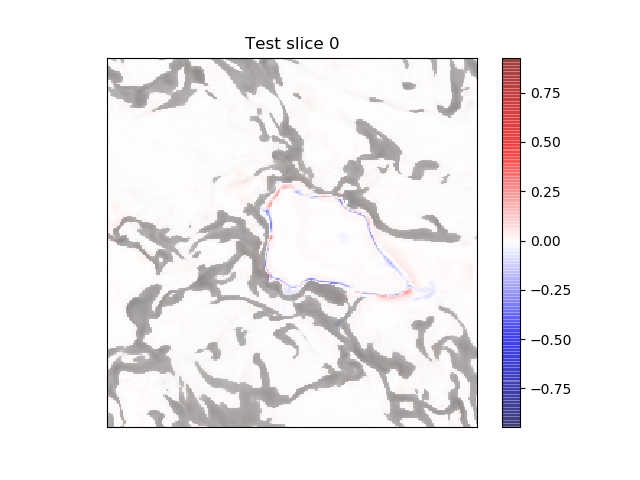}
\includegraphics[width=2in,trim={1in .4in .6in .55in}, clip]{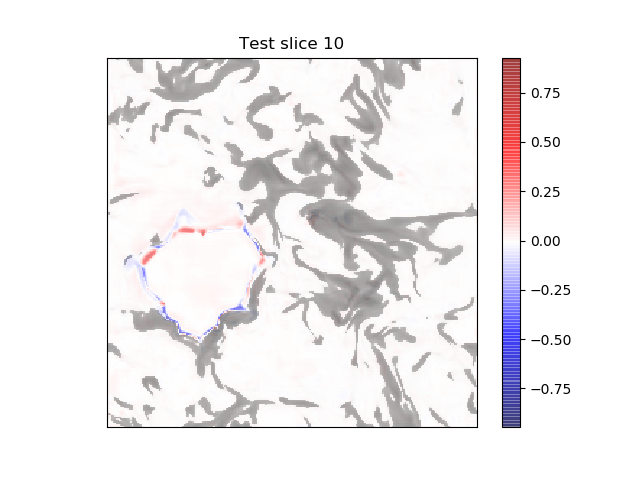}
\includegraphics[width=2in,trim={1in .4in .6in .55in}, clip]{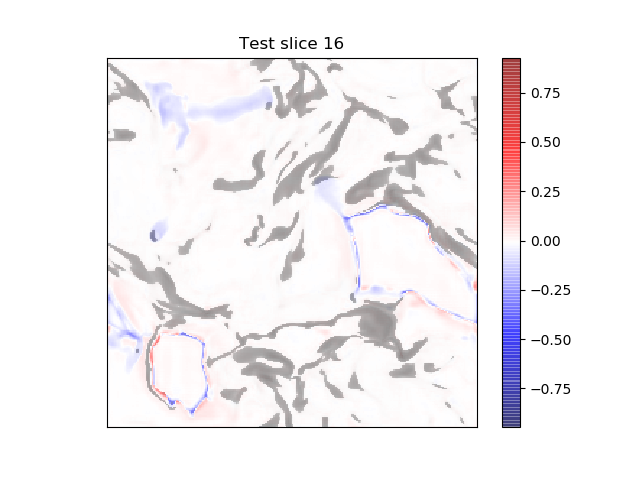} \\
\vspace{0.1in}
\includegraphics[width=2in,trim={1in .4in .6in .55in}, clip]{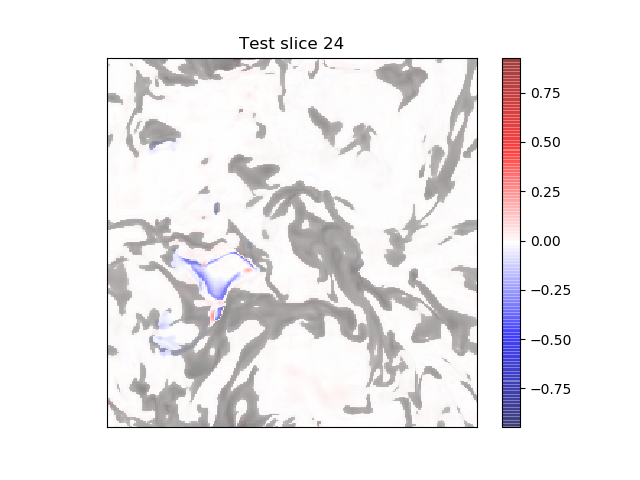}
\includegraphics[width=2in,trim={1in .4in .6in .55in}, clip]{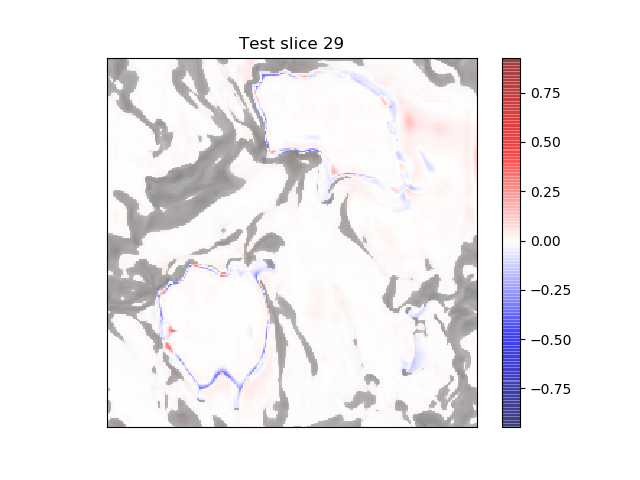}
\includegraphics[width=2in,trim={1in .4in .6in .55in}, clip]{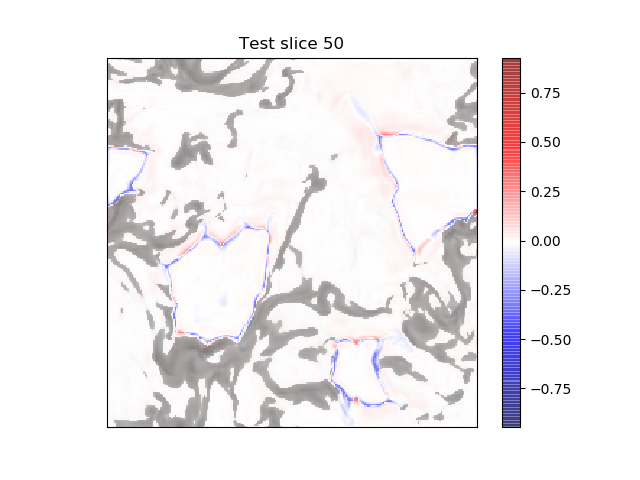} \\
\caption{
Example residuals from a Residual U-Net trained on the density regression task using the mean squared error loss function.
Positive residuals, shown in shades of red, correspond to over-estimation, while negative residuals, shown in shades of blue, correspond to under-estimation.
As with Figure \ref{fig.density_segmentation}, the side length of each tile is 5 pc and the gray scale components represent re-scaled density values.
\label{fig.regress_density_examples} 
}
\end{figure*}

Figure \ref{fig.regress_density_examples} displays the example prediction residuals for several samples from the test set.
Note that the larger residuals tend to be clustered together near the edges of structures, similar to what was observed in the density segmentation setting.

\begin{table*}
\centering
\begin{tabular}{|l l|rr|}
\hline
Performance Stat. & Distribution Stat. & Density Regression & $^{12}$CO Regression \\ \hline
Mean              & Mean               & -0.0527            & -0.019               \\
          & Std. Error         & 0.0009             & 0.0008               \\
          & 85\% Conf. Int.    & (-0.054, -0.0513)  & (-0.0201, -0.0179)   \\ \hline
Std. Dev.         & Mean               & 0.2012             & 0.3483               \\
          & Std. Error         & 0.0031             & 0.0011               \\
          & 85\% Conf. Int.    & (0.1968, 0.2058)   & (0.3466, 0.3499)     \\ \hline
Skew              & Mean               & -3.8254            & -13.17               \\
          & Std. Error         & 0.0211             & 0.0854               \\
          & 85\% Conf. Int.    & (-3.8562, -3.7946) & (-13.2945, -13.0454) \\ \hline
Score             & Mean               & -4.0793            & -13.5372             \\
          & Std. Error         & 0.01778            & 0.0854               \\
          & 85\% Conf. Int.    & (-4.1053, -4.0534) & (-13.6618, -13.4126) \\ \hline
\end{tabular}
\caption{
Regression task performance statistics collected by training and evaluating 60 randomly initialized networks on the same training, validation, and testing splits.
The first column indicates a statistic that was computed over the residuals of each trained network, while the second column indicates a statistic that was applied to the results of the column one statistic.
The fourth element of the first column, Score, refers to the regression score defined in Section \ref{sec.regression}.
CASI is able to reliably obtain a residual distribution with a mean near zero and a small standard deviation, indicating a tight residual distribution that is clustered about the origin.
For both tasks the mean and skew components feature negative values, indicating that CASI tends to under-estimate values more often than it over-estimates values.
Additionally, the negative skew value indicates that the tail of the residual distribution is longer in the negative direction, thus the largest errors tend to be under-predictions.
However, the fact that the residual distribution is tightly grouped about the origin indicates that the relatively large skew value is not concerning and due in part to the characteristics of the input data.
}
\label{table.regression_scores}
\end{table*}

\subsection{Case Study 2: Synthetic Molecular Emission}
The $^{12}$CO data features position-position-velocity coordinates, rather than the position-position-position coordinates used for the density data.
In both the segmentation and regression tasks the input data is inspected along the velocity axis such that the network is provided with position-position slices, those slices are divided into training, validation, and testing splits, then each data split is normalized by subtracting the mean and dividing by the standard deviation.

\subsubsection{CO Segmentation}
The U-Net attains slightly lower performance in the $^{12}$CO tasks, when compared with corresponding density tasks, even though the training set is more than a factor of two larger.
This indicates that the relationship between the $^{12}$CO observations and the constructed tracer data may be more complex than the relationship between the density data and corresponding tracer. 

\begin{figure*}
\centering
\includegraphics[width=2in,trim={1in .4in 1in .5in}, clip]{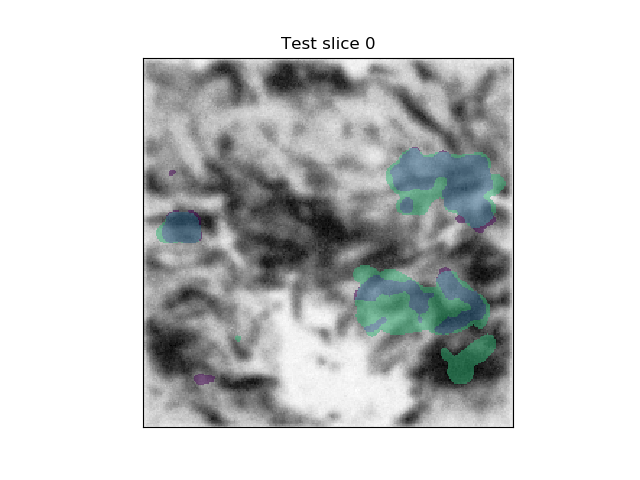}
\includegraphics[width=2in,trim={1in .4in 1in .5in}, clip]{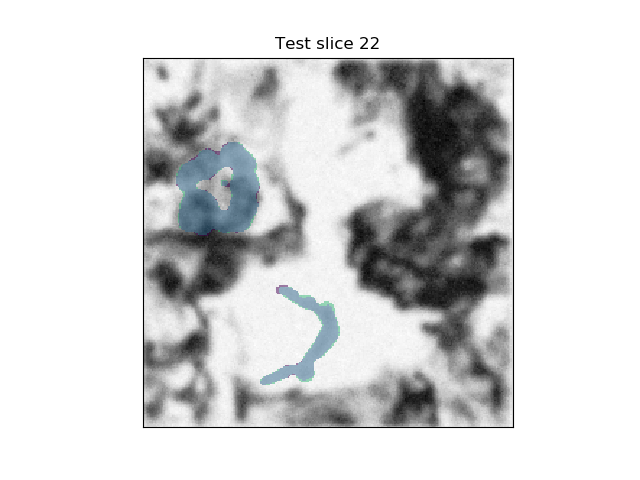}
\includegraphics[width=2in,trim={1in .4in 1in .5in}, clip]{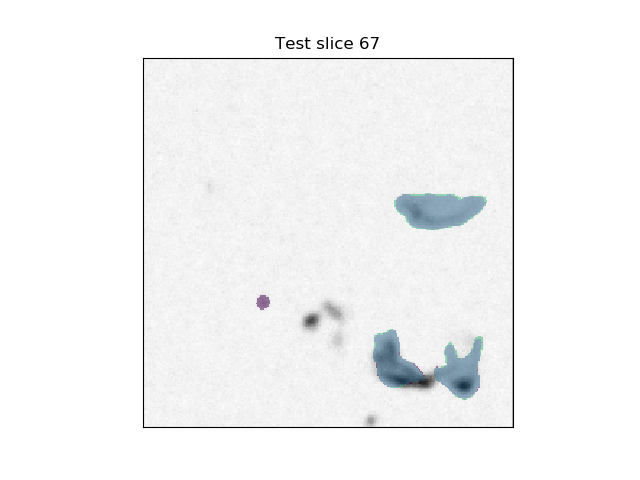}\\
\vspace{0.1in}
\includegraphics[width=2in,trim={1in .4in 1in .5in}, clip]{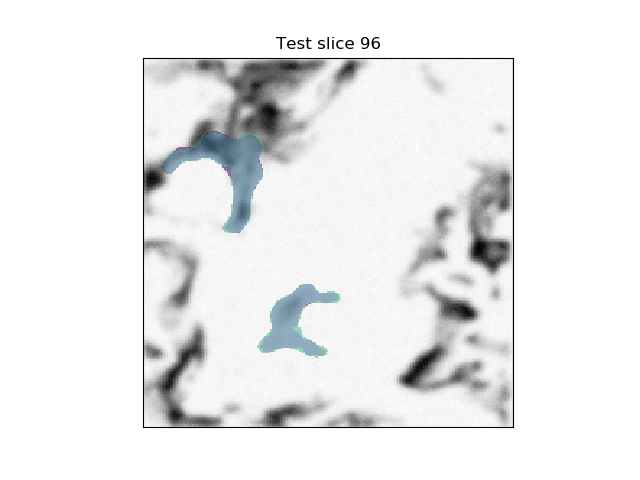}
\includegraphics[width=2in,trim={1in .4in 1in .5in}, clip]{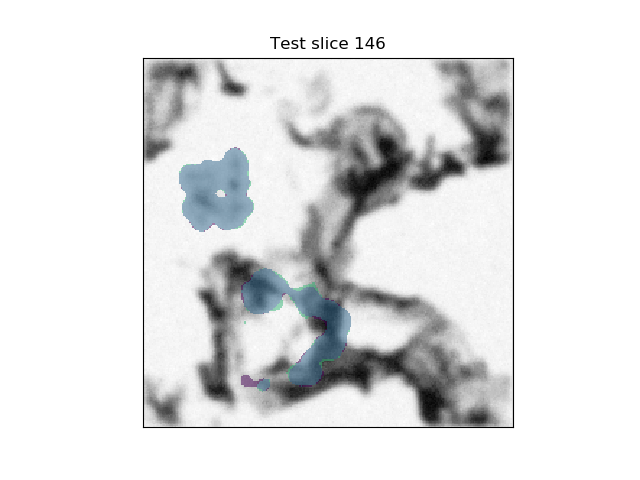}
\includegraphics[width=2in,trim={1in .4in 1in .5in}, clip]{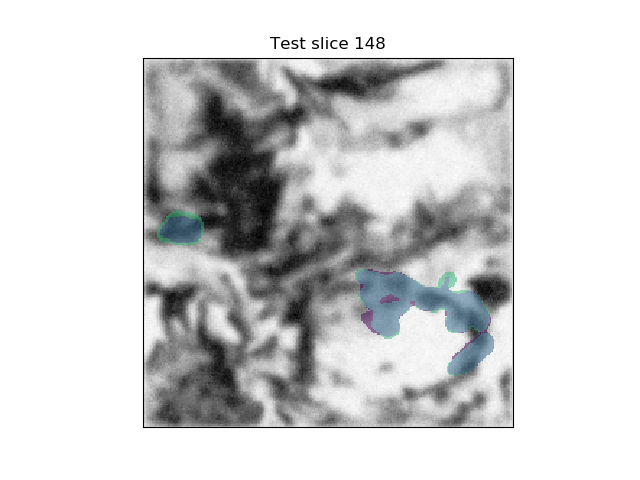}
\caption{
$^{12}$CO segmentation predictions from a Residual U-Net on samples randomly selected from the test set.
As with Figure \ref{fig.density_segmentation}, the side length of each tile is 5 pc, the gray scale components represent re-scaled density values, true positives are shown in blue, false positives are shown in purple, false negatives are shown in green, and true negatives are not displayed.
}
\label{fig.classify_co_diffs}
\end{figure*}

Figure \ref{fig.classify_co_diffs} shows example predictions, which feature similar characteristics to the density segmentation predictions.
The major structures are all correctly identified, with some smaller structures being missed, and errors clustered along the edges of larger structures.

\begin{figure}
\centering
\includegraphics[width=3.5in,trim={0in 0in 0in .55in}, clip]{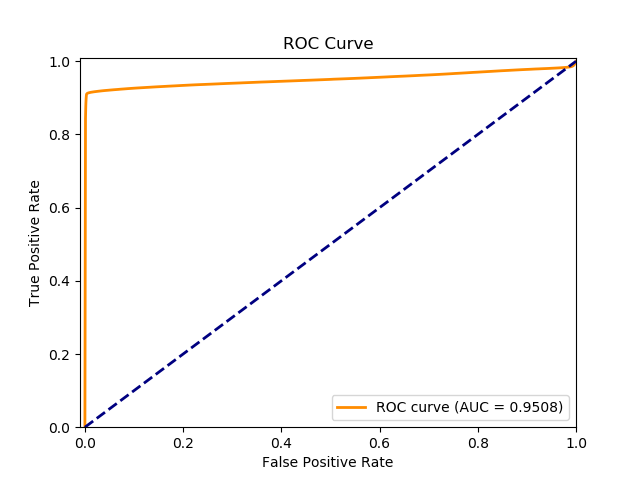}
\caption{
Example ROC curve for a Residual U-Net trained on the $^{12}$CO segmentation task.
The dashed blue line represents $y=x$, which corresponds with the expected performance of a random binary classifier.
A true positive rate of 91.45\% is obtainable with a false positive rate of 1\%, supporting the proposal that this method may perform well as a content filter.
\label{fig.classify_co_roc}
}
\end{figure}

The ROC curve, provided in Figure \ref{fig.classify_co_roc}, features a sharp curve that is pushed up towards the upper-left corner of the plot, where the model reaches a true positive rate of 91.45\% while maintaining a false positive rate of 1\%.
This accuracy is slightly lower than that achieved by the density segmentation task, however, the results still constitute excellent performance.

An investigation of final model performance variation due to random initialization is provided in column two of Table \ref{table.segmentation_scores}, which shows that CASI is robust to random initialization on the CO segmentation task.

\subsubsection{CO Regression}
For the $^{12}$CO regression task an element-wise logarithm operation is applied prior to the normalization operations in order to further reduce the dynamic range of the data.
Specifically we modify the data using
$$
x' = \ln(1 + x - min(x)),
$$
where $x$ is one of the training, validation, or testing sets, and $min(x)$ is the element-wise minimum.
Subtracting by the minimum value and adding 1 ensures that there are no invalid output values and that all values fall within the compressive regime of the logarithm.

We investigated the effect of random initialization on this task using the same experiment structure seen for the other conditions.
These results are reported in column two of Table \ref{table.regression_scores}.

\section{Conclusions} \label{conclusions}
Our results indicate that methods from deep learning, namely the U-Net and its variants, are a flexible and effective tool for learning relationships in simulated density and $^{12}$CO data.
Moreover, our algorithm is completely general and could be trained to identify other astronomical signatures, such as protostellar outflows, filaments and dense cores, given appropriate training sets.

Our models learn well under several different conditions and generalize to unseen data from the same distribution with a minimal performance impact.
Additionally, CASI features a low false positive rate and a clustering of errors that makes it well-suited to assisting astronomers by filtering large-scale survey data that is being inspected by humans.

We also note that \textsc{CASI} is relatively quick to train, especially on smaller datasets, taking approximately 2 seconds per epoch for the Density tasks (approximately 8 minutes for 200 epochs) and 7 seconds per epoch for the CO tasks (approximately 25 minutes for 200 epochs).
After training, \textsc{CASI} can process more than 100 samples per second, allowing for rapid application to new data.
With fast training times and even faster post-training predictions, \textsc{CASI} may be rapidly applied to new datasets with minimal overhead.\footnote{The computer that was used to collect timing information was outfitted with an Intel i7-6700K CPU and a Nvidia GTX 1080Ti GPU.}

Despite the generally positive results presented, there are several important research directions surrounding the application of deep learning techniques to facilitate the analysis of astronomical image data that have not been addressed. 

First, all results presented in this work focus upon learning from simulated data, but in order to assist in the processing of survey data these models must operate on true observations that may greatly differ from the simulated data that they were trained upon.
For example, we adopt a simple CO abundance model and do not take into account chemistry.
\citet{boyden18} show that self-consistently computing abundances and temperatures can produce statistically different emission maps.
However, \citet{xu17} demonstrate that synthetic dust emission maps of the simulations also utilized here can be used to successfully train a random forest algorithm to correctly identify observed bubbles.
This lends confidence that our CO emission maps have, at minimum, similar underlying morphologies to observational data.
We extend our study to observational data in Xu et al. (in prep) and demonstrate that training sets based on synthetic CO emission can indeed be applied to observed CO data.
Beyond assessing and improving the simulations that are used to generate training data, a comprehensive investigation of regularization and data augmentation techniques may lead to models that are better able to bridge the gap between simulation and observations.

Second, our methods leverage the high fidelity information and annotations provided by the simulations to learn relationships in a supervised setting.
However, there exist considerable amounts of unlabeled survey and observational data that may be utilized in semi-supervised or unsupervised approaches.
Semi-supervised and unsupervised approaches could reduce or remove the overhead involved with hand labeling and curating large data sets, while still drawing insights from said data.

Finally, only 2D models were investigated in this work, which ignore the 3D structure present in density and $^{12}$CO cubes.
We have found that 2D models seem to be sufficient for solving certain problems in this domain, certainly the benchmarks investigated here are well solved by 2D models, but some problems may require models with greater knowledge of 3D structure.

3D convolutional models have begun to find application in human action recognition \citep{ji20133d}, object detection in 3D point clouds \citep{Maturana-2015-6018}, medical imaging \citep{khosla20183d}, and other domains \citep{tran2015learning}.
These 3D models may also be well-suited to identifying structures in stellar feedback, and we begin to explore such models, as well as their application to observational data, in upcoming work (Xu et al.~in prep).

\acknowledgments
CVO, DX, SSRO, and RAG were supported by NSF grant AST-1812747. SSRO also acknowledges support from NSF Career grant AST-1650486.

\software{Astropy, Keras, Matplotlib, Numpy, Pandas, Python, Scikit-learn, Tensorflow,}

\bibliographystyle{yahapj}
\bibliography{main}

\appendix
\section{Relevant Neural Network Operations}\label{appendix.nn-ops}

\subsection{Batch Normalization}
Batch normalization allows a network to re-normalize data at arbitrary points during the forward pass using moving mean and standard deviation statistics calculated over training batches.
Following the description of batch normalization provided by \cite{ioffe2015batch}, if $\mathcal{B} = \{ x_1, x_2, ..., x_n \}$ represents a batch of training samples then the mean and variance of the batch are calculated as
\begin{align*}
&\mu_{\mathcal{B}} = \dfrac{1}{n} \sum_{i=1}^{n} x_i,
&& \sigma_{\mathcal{B}}^2 = \dfrac{1}{n} \sum_{i=1}^{n} (x_i - \mu)^2.
\end{align*}
The data are then normalized using the batch mean and variance
$$
\hat{x}_i = \dfrac{x_i - \mu_{\mathcal{B}}}{\sqrt{\sigma_{\mathcal{B}}^2 + \epsilon}},
$$
where $\epsilon$ is an arbitrary constant used for numerical stability.
Finally, the output of the batch normalization is calculated using
$$
y_i = \gamma \hat{x}_i + \beta,
$$
where $\gamma$ and $\beta$ are learned parameters that allow the network to reverse or modify the batch normalization procedure when beneficial.

Batch normalization is applied slightly differently during training and inference, though this is handled internally by most deep learning frameworks.
Interested readers should refer to \cite{ioffe2015batch}.

\subsection{Convolution}\label{appendix.conv}

A 2D convolution in this context involves an image with dimensions (image height, image width, image channels), or $(H_i,\ W_i,\ C_i)$, and a set of filters with dimensions (filter count, image channels, filter height, filter width), or $(F,\ C_i,\ H_f,\ W_f)$.
The convolution is computed by sliding each filter over the spatial dimensions of the image.
At each location an element-wise product between the filter and the corresponding image pixels is computed, the results of which are summed and become a single pixel in the output of the convolution.
The sliding behavior of the convolution is controlled by horizontal and vertical stride parameters, $s_h$ and $s_v$, which indicate how far the filter should move in each direction after each calculation.
The output of the convolution described above would have the dimensions
$$
\left( \frac{H_i - H_f + 1}{s_v},\ \frac{W_i - W_f + 1}{s_h},\ F \right).
$$

It is common to pad the image with zeros in order to force the dimensions of the output into desired values.
Notably, if the spatial dimensions of the filter are odd and the image is padded by $\lfloor H_f / 2 \rfloor$ on the top/bottom and $\lfloor W_f / 2 \rfloor$ on the left/right then the the output of the convolution will have the dimensions $(H_i,\ W_i,\ F)$.
This is referred to as the ``same'' padding scheme, since the output has identical spatial dimensions to the input.
In practice, this operation is usually applied to a batch of several images in parallel.

\subsection{Max Pooling} \label{appendix.pooling}

The max pooling operation is designed to reduce the spatial dimensions of an image while keeping the most important data intact.
It does this by inspecting small sub-regions of the image, commonly $2\times 2$ windows, and filtering out the maximum value in that sub-region.
The max pooling operation, like the convolution described above, has stride parameters which adjust the spatial relationship between the sub-regions.
It is common to have strides that are equal to the size of the sub-regions, resulting in disjoint sub-regions which fully cover the input image.

Note that max pooling is applied to each channel independently, and thus the result of applying a max pooling operation with a 2x2 window and a stride of 2 to a (64, 64, 3) image would be a (32, 32, 3) image.

\begin{figure}
\centering
\includegraphics[width=3.5in]{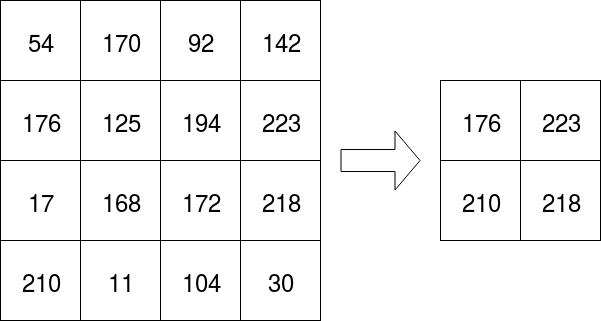}
\caption{Max pooling with a $2 \times 2$ window, used to map a $4 \times 4$ input to a $2 \times 2$ output. \label{fig.maxpool}}
\end{figure}

\subsection{Nearest-Neighbor Interpolation}

Nearest-neighbor interpolation is an extremely simple up-sampling operation that increases the spatial dimensions of an image by an integer factor, $n$, by expanding each pixel into an $n \times n$ block with identical values.
This may be used to reverse the effects of a max pooling operation, though some detail is lost.

\begin{figure}
\centering
\includegraphics[width=3.5in]{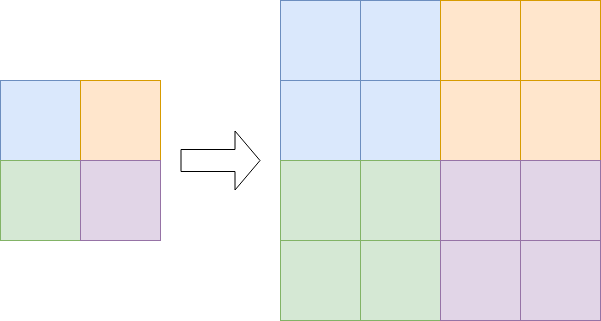}
\caption{
Nearest-neighbor interpolation with a $2 \times 2$ window, used to map a $2 \times 2$ input to a $4 \times 4$ output.
\label{fig.upsampling}}
\end{figure}

\subsection{Activation: Exponential Linear Units}

Introduced by \cite{clevert2015fast}, the exponential linear activation function is defined as:
\begin{align*}
\text{ELU}(x) =
\begin{cases}
x & \text{if } x \geq 0 \\
\alpha (e^x -1) & \text{if } x < 0, \\
\end{cases}
\end{align*}
where $\alpha$ controls the negative saturation value of the function.
Use of exponential linear units (ELU) has been shown empirically to allow faster and more robust training of deep neural networks when compared to rectified linear units (ReLU) and other common activation functions.

Scaled Exponential Linear Units (SELU), defined as 
\begin{align*}
\text{SELU}(x) = \lambda
\begin{cases}
x & \text{if } x \geq 0 \\
\alpha (e^x -1) & \text{if } x < 0, \\
\end{cases}
\qquad \lambda > 1,
\end{align*}
exhibit similar properties to ELUs but with the added benefit of having a normalizing effect on network activations, similar to batch normalization.
See \cite{klambauer2017self} for more details.

\subsection{Residual Connections} \label{appendix.residual}
Sometimes referred to as skip connections, this architecture component can improve performance \citep{he2016deep}, reduce training instability in deeper networks \citep{he2016deep}, and encourage iterative inference \citep{jastrzebski2017residual}.

\begin{figure}
\centering
\includegraphics[width=3in]{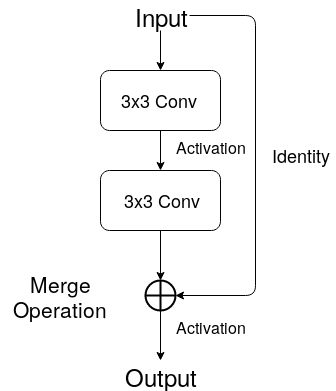}
\includegraphics[width=3in]{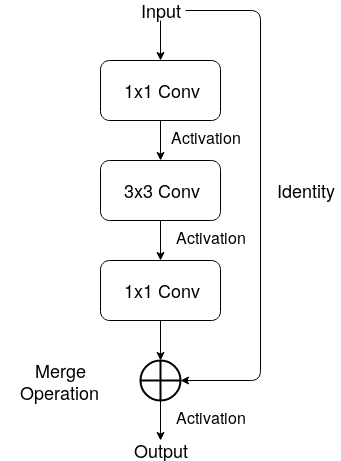}
\caption{
Left: A basic residual block using $3 \times 3$ filters, the number of filters used in each convolution is a free parameter that must be selected.
Common activation functions include ReLU, sigmoid, and tanh.
Common merge operations include concatenation, element-wise addition, and element-wise maximum \citep[maxout,][]{goodfellow2013maxout}.
If addition is used as the merging operation then a projection skip-connection, commonly implemented using a $1 \times 1$ convolution, may be required in place of the identity skip-connection in order to obtain the correct dimensions for the merge operation.
Right: A bottleneck residual block, which uses $1 \times 1$ convolutions in order to reduce the number of parameters required, relative to the basic residual block.
If the input volume has $n$ channels it is common to use $n/2$ or $n /4$ filters in the first two convolutions followed by $n$ channels in the final convolution, this compresses the data before the larger convolution is applied  resulting in a reduced number of parameters.
\label{fig.residual}}
\end{figure}
\end{document}